\documentclass[prd,twocolumn,showpacs,preprintnumbers,floats,floatfix,apsrev,amsmath,amsfonts,nofootinbib]{revtex4-1}
\usepackage{xcolor}
\usepackage{graphicx}
\usepackage{dcolumn}
\usepackage{txfonts}
\usepackage{multirow}
\usepackage{subcaption}
\usepackage{natbib}
\usepackage{hyperref}
\usepackage{hypernat}
\usepackage{verbatim}
\usepackage{physics}
\usepackage{amsmath}

\newcommand{\lp}{\left(}
\newcommand{\rp}{\right)}
\newcommand{\lc}{\left[}
\newcommand{\rc}{\right]}

\def\be{\begin{equation}}
\def\ee{\end{equation}}
\def\ba{\begin{eqnarray}}
\def\ea{\end{eqnarray}}

\def\b{\beta}
\def\d{\delta}

\def\vf{\varphi}
\def\g{\gamma}
\def\h{\eta}

\def\k{\kappa}

\def\r{\rho}

\def\s{\sigma}

\def\z{\zeta}

\def\ci{{\cal I}}

\def\cl{{\cal L}}

\def\co{{\cal O}}
\def\cp{{\cal P}}

\newcommand{\mpl}{M_{\mathrm{Pl}}}

\newcommand{\fnl}{f_{\mathrm{NL}}}
\newcommand{\fnlo}{f_{\mathrm{NL}}^{\mathrm{local}}}
\newcommand{\gnl}{g_{\rm NL}}

\newcommand{\ns}{n_{s}}

\newcommand{\rec}{r_{\rm dec}}
\newcommand{\sosc}{\sigma_{\rm osc}}
\newcommand{\sk}{\sigma_{\rm k}}
\newcommand{\sq}{\sigma_{\rm q}}
\newcommand{\smax}{\sigma_{\rm max}}
\newcommand{\fosc}{f_{\rm osc}}

\begin{document}

\title{Viable Curvaton Models from the $f_{NL}$ Parameter}

\author{L.~F.~Guimar\~aes${}^{1,2}$}
\email{lfog@cbpf.br}

\author{F.~T.~Falciano${}^{1,3}$}
\email{ftovar@cbpf.br}

\affiliation{
${}^1$CBPF - Centro Brasileiro de Pesquisas F\'isicas, Xavier Sigaud st.\ 150, zip 22290-180, 
Rio de Janeiro, RJ, Brazil \\
${}^2$Dipartimento di Fisica, Università di Pisa and INFN, Sezione di Pisa, Largo Pontecorvo 3, I-56127 Pisa, Italy \\
${}^3$PPGCosmo, CCE - Universidade Federal do Esp\'irito Santo,
zip 29075-910, Vit\'oria, ES, Brazil}

\date{\today}

\begin{abstract}
We show how to build a curvaton inflationary model motivated by scale-dependent non-Gaussianities of cosmological perturbations. In particular, we study the change of sign in the $\fnl$ parameter as a function of the curvaton field value at horizon crossing and identify it with the cosmic microwave background pivot scale. We devise a procedure to recover the curvaton model that provides the desired $\fnl$ parameter. We then present a concrete example of $\fnl$ and construct its parent model. We study the constraints applied to this model based on considerations taken on $\fnl$. We show that the hemispherical asymmetry can also be used to constrain the scale-dependence of $\fnl$ and the model parameters.
\end{abstract}



\maketitle

\section{Introduction}

Modern cosmology is based on a six parameter model of the early universe that has been tested with great precision by measuring the Cosmic Microwave Background radiation (CMB) anisotropies. The latest Planck results~\cite{Planck_2018_1, Planck_2018_6, Planck2018inflation} confirm the concordance model and provides accurate information on the cosmological parameters. In particular, the primordial density perturbations are consistent with gaussian curvature perturbations, and the data is well suited to the inflationary scenario~\cite{Martin2014}. The power spectrum has a scalar spectral index of $n_s = 0.9649 \pm 0.0042$ and a small tensor-to-scalar ratio $r < 0.064$. The primordial non-Gaussianities have not yet been detected but there are constraints for all types of shapes. In particular, the local shape is constrained to be $\fnlo = -0.9 \pm 5.1$ \cite{Planck_2018_11}.

Inflation offers a mechanism to explain the existence of the primordial cosmological perturbations. The most simple scenario is the single-field inflation (SFI) where a scalar field follows a slow-roll dynamics and is simultaneously responsible to drive the almost exponential expansion of the universe and its density perturbations work as seeds for the CMB~\cite{Lyth_1998}. Thus, a slow-roll SFI models provides the observed Gaussian and almost scale-independent temperature fluctuations~\cite{Creminelli_2004}.

There are, however, at least two motivations to consider alternatives to SFI. In recent years, theoretical arguments indicate that the scenario might be not in the string-theory landscape but in its swampland~\cite{Obied_2018, Agrawal_2018, Garg_2019, Bedroya_2020}. In addition, SFI is not suitable to address the CMB anomalies or observation of primordial non-Gaussianities. The CMB anomalies manifested on the Planck data~\cite{Planck_2015_16} still have a low statistical significance of $3\sigma$. Nevertheless, the fact that they were measured by two different surveys, namely WMAP and then Planck satellites, suggest that these anomalies might not be just a systematic error or foreground contamination, and if they exist, the statistical anomalies go against the cosmological principle~\cite{Copi_2010, Schwarz_2016}.

SFI is not the unique successful scenario of the early universe. There are mainly two alternative classes models. One, are the bouncing models, which also give an almost scale invariant power spectrum with the correct redshift tilt and negligible production of gravitational waves~\cite{Guimaraes_2019, Wands_1999, Falciano2008, PintoNeto:2012ug, Falciano:2013uaa, Vitenti:2012cx, Peter:2015zaa, Peter:2008qz, Brandenberger:2016vhg, Lilley:2015ksa}. Many bouncing models avoid constraints coming from the Swampland conjectures, specially by the absence of a de Sitter expansion phase~\cite{Brandenberger_2011, Laliberte_2020, Brandenberger_2020}. Another route is to preserve an inflationary phase, albeit not in a single-field slow-roll setting. Multi-Field Inflation (MFI) models~\cite{Wands_2007}, in which more than one scalar field ruling the inflationary regime, could be a requirement for inflation to happen in the string theory landscape~\cite{Achucarro_2019}, while Warm Inflation provides alternative routes around the issues~\cite{Motaharfar_2019, Das_2019}. 

Among the MFI models, the curvaton models are simple extensions of SFI with the addition of only one extra scalar field~\cite{Lyth_2002, Lyth_2003}. In this scenario, the background dynamics is still driven by the inflaton but the cosmological perturbations now come from the density fluctuations of the curvaton field. A distinct feature of these models is that the curvaton produces isocurvature perturbations instead of the common inflaton adiabatic perturbations. Only after its own decay into radiation that the curvaton isocurvature modes turn into adiabatic, which then seeds the CMB. This scenario alleviates the constraints on the inflaton field~\cite{Vennin_2015} while still producing the observed almost scale-invariant spectrum and the negligible amplitude of the primordial gravitational waves. Another advantage of this scenario is that it allows for large non-Gaussianities, and indeed, much higher than in SFI.

A possible mechanism to account for the CMB anomalies is to consider non-gaussian super-Hubble perturbations. A non-gaussian mixing of long and short scale modes~\cite{Schmidt_Hui_PRL} breaks the perturbations' isotropy and can explain the hemispherical asymmetry~\cite{Byrnes_Tarrant_2015, Byrnes_2016, Byrnes_2016_2, Kenton_2015}. In particular, the curvaton scenario predicts non-gaussian primordial perturbations due to a quadratic dependence on the curvaton field in $\zeta$~\cite{Lyth_2002}.

The self-interacting models~\cite{Huang_2008, Enqvist_2010, Byrnes_2011} are of particular interest inasmuch they provide scale-dependent non-Gaussianities that allows for the non-Gaussianity parameters $\fnl$ and $\gnl$ to vary orders of magnitude between different scales. In this case, it is possible to have large non-Gaussianity at large scales and still satisfy the observational constraint on $\fnl$. In addition, scale-dependence models can also modulate the non-Gaussianities and get the right amount of power asymmetry in the CMB\cite{Schmidt_Hui_PRL, Adhikari_2018, Hansen_2018}.

In the present work, we show how to construct viable curvaton models from the properties of the $\fnl$ parameter. In particular, due to the change of sign, we manage to have a $\fnl$ close to zero at the observable scales but still have large non-Gaussianities away from the pivot scale. The paper is organized as follow. In sec.~\ref{SecCurv} we review the self-interacting curvaton scenario and in section~\ref{SecRecon} we show how to construct curvaton models that implement the desired features of $\fnl$. In sec.~\ref{SecLinear} we analyze the parameter space of one of such models and show that our procedure alleviates the fine-tuning of such models. In sec~\ref{Con} we conclude with final remarks. Throughout the paper, unless explicitly written, we use Planck mass $\mpl=1$.

\section{Curvaton scenario} \label{SecCurv}
\subsection{Self-interacting curvaton scenario} \label{SubSec_SI_Curvaton}

The curvaton scenario goes beyond SFI by the addition of a second scalar field dubbed the curvaton. Usually, this extra scalar field is minimally coupled to gravity and does not interact with the inflaton. The latter drives the background dynamics while the curvaton produces the observed cosmological perturbations. There are also interactive models~\cite{Langlois_2004, Ichikawa_2008} where the potential has a cross term coupling the curvaton with the inflaton. These interactive models satisfy the observational constraints but at the cost of increasing the number of free parameters of the model. Here, we shall consider only self-interacting curvaton models, which have scale-dependent non-Gaussianity~\cite{Byrnes_Tarrant_2015}, such that the Lagrangian reads

\begin{align}
\cl (\vf,\s)=
K(\vf) + K(\s) + V(\vf) + V(\s)\ , 
\end{align}
where $K(X)$ and $V(X)$ denotes the kinetic and potential terms of the inflaton and curvaton fields, respectively. In contrast to SFI models, in the curvaton scenario, the inflaton has a negligible contribution to the cosmological perturbations due to a lower inflaton mass $m_\varphi$ as compared to the SFI models~\cite{Byrnes_2014}~\footnote{The scenario actually allows for both fields to contribute to cosmological perturbations~\cite{Langlois_2004, Ichikawa_2008}}. As a consequence, the magnitude of tensor perturbations is likewise negligible as compared to the SFI. Notwithstanding, the energy density of the curvaton is always sub-dominant and do not contribute to the background dynamics. It is the inflaton slow-roll regime that drives the almost exponential expansion of the universe, while the curvaton follows its own evolution, which does not need to the frozen but can be a slow-roll different from the inflaton dynamics. 

As usual, reheating takes place at the end of the inflaton slow-roll regime, when it oscillates around the minimum of the potential with an equation of state $p =\omega \rho$ with $\omega \approx 0$. During this process the inflaton decays into radiation. After decay we are left with a reheated universe, with energy density radiation-dominated.

In our scenario, the curvaton field follows a similar decay regime, albeit delayed in time. Thus, we consider potentials for the curvaton with a local minimum, that can be approximated by a quadratic potential, and where the coherent oscillations makes the curvaton decays as pressureless dust~\footnote{There are models in which the behavior of the potential at small values of the field is not quadratic, see \cite{Kawasaki_2011} and references herein.}. In addition, we assume the sudden decay approximation in which the curvaton instantaneously decays into radiation when its decay rate equals the Hubble parameter, i.e. $\Gamma_{\sigma} = H$~\footnote{It can be shown that the sudden decay is a good approximation for the exact gradual decay. Moreover, it does not impact on the primordial observables~\cite{Meyers_Tarrant}.}. 

During the inflationary phase the curvaton produces only isocurvature perturbations. Due to thermal and chemical equilibrium, after the curvaton decay they are then converted into adiabatic perturbations. This conversion process was first proposed by Mollerach~\cite{Mollerach_1990} and latter applied to the curvaton scenario by \cite{Lyth_2002, Lyth_2003}. The transfer of isocurvature perturbation into curvature perturbation can be described as~\cite{Lyth_2002}

\ba \label{zcurvaton}
\z \sim \rec \d\ ,
\ea

where $\rec$ and $\d$ are respectively the curvaton fractional energy density and isocurvature perturbation and $\z$ is the final adiabatic perturbation. The fractional energy density gives the curvaton contribution to the total energy density and reads

\ba \label{def_r}
\rec = \left. \frac{3\r_{\s}}{3 \r_{\s} + 4\r_{\g}}\right |_{\rm dec} \sim \frac{V(\s_{\mathrm{dec.}})}{3 \Gamma^2}\ ,
\ea

where $\r_{\s}$ and $\r_{\g}$ are respectively the curvaton and the radiation density at the time of the curvaton decay. During its reheating, the curvaton redshifts slower than radiation, since it behaves as dust, hence if the curvaton decays long after the inflaton, the curvaton dominates the energy density of the universe and $\rec \sim 1$. On the other hand, if the decay happens shortly after the inflaton's decay, then $\rec \ll 1 $, which means large non-Gaussianity (see discussion after eq.~\eqref{foscdef}). Therefore, we assume that the curvaton decays not long after the inflaton and hence $\rec \sim 10^{-2}$.

We can calculate the curvature perturbation by the $\delta N$ formalism~\cite{Lyth_2005_PRL}. The difference in the curvature perturbations equals the number of e-folds between the two hypersurfaces $\z \equiv \d N $. For single source curvaton models hence we have 


\begin{align}
\z  
&\approx N_{,\s} \d \s  
+\frac{1}{2} N_{,\s \s} \d \s^2  + \cdots 
\nonumber
\end{align}

where $N_{,\s}$ is the derivative of $N$ with respect to the curvaton field $\s$ at the initial hypersurface. The curvature perturbation power spectrum is defined as

\begin{align}
\cp_{\z}({k}) &=  {N_{,\s}}^{2} \, \cp_{s} 
\approx \frac{\rec^2}{9 \pi^2} \left(\frac{\sosc'}{\sosc}\right)^2 H_{k}^{2} \label{pert_amplitude}
\end{align}

where $\cp_{s}({k})=H_{k}^{2}/(4\pi^2)$ is the scalar power spectrum for the mode $k$. We have that $\sigma_{\rm osc} = \sigma_{\rm osc}(\s_k)$ is the amplitude of the oscillations~\footnote{We prove that $\sosc$ depends on $\sk$ later in this same section.}. One can show that $N_{,\s} = \frac23\rec \, (\sosc'/\sosc)$. Therefore, the spectral index reads

\begin{align} \label{nscurvaton}
\ns - 1 &\approx 2 \frac{\dot{H}_{k}^2}{H_{k}^{2}} + 2 \frac{V_{,\s \s}}{3 H_{k}^{2}} \approx -2\epsilon_{H} + 2\h_{\s}\quad ,
\end{align}

Notwithstanding the inflaton still gives important contributions since it dominates the background dynamics.  In the above equation, we defined the slow-roll parameters as usual, namely, 

\begin{align}
\epsilon_{H} \equiv \frac{\dot{H}_{k}^2}{H_{k}^{2}} \quad  ,  \quad 
\eta_{\s} \equiv  \frac{V_{,\s \s}(t_k)}{3 H_{k}^{2}}
\end{align}

Even though the curvaton does not interact with the inflaton, the scalar perturbations and spectral index have contributions from both fields. The spectral index \eqref{nscurvaton} has a leading contribution $\epsilon_{H}$, the inflaton slow-roll parameter. The curvaton $\eta_{\s}$, if positive, must be sub-leading and of order $10^{-2}$ or lower so that the spectrum is red and quasi-scale invariant~\footnote{Models with negative spectral index would be preferred because of the requirement of a red spectrum, and would alleviate some conditions imposed on the inflaton, via $\epsilon_{H}$}. The tensor-to-scalar ratio $r$ is largely suppressed in the curvaton scenario as compared to SFI, 

\begin{align}
r = 16 \epsilon_{H}  \frac{\cp_{\vf}}{\cp_{\z}} \approx 0 \ll r_{SFI}\quad ,
\end{align}


where again we are dealing with the fact that the inflaton does not contribute to the perturbations $\cp_{\z} \gg \cp_{\vf}$. One of the advantages of the curvaton scenario is to evade the need for $\epsilon_{H} \propto 1/N^{2}$ (for $N$ around 60 e-folds) in order to fit the current observational sensibility, $r < 10^{-2}$~\cite{Vennin_2015}. Indeed, SFI models that lead to $\epsilon_{H} \propto 1/N$, such as chaotic inflation~\cite{Linde_1983, Lyth_1998}, can now be used as the inflaton component of the curvaton scenario, since they satisfy both constraints on $r$ and $n_{s} - 1$.

In order to quantify the amount of non-Gaussianity in the model, we can be Taylor expand the curvature perturbation $\zeta(k)$ in terms of its Gaussian component $\z_{G}$ as~\cite{Sasaki_2006}

\ba \label{z3o}
\z = \z_{G} + \frac{3}{5}\fnl\z_{G}^{2} + \frac{9}{25}\gnl\z_{G}^{3} + \co(\z_{G}^{4}) \ .
\ea

By definition, the non-linearity parameters $\fnl$ and $\gnl$ encode, respectively, the non-Gaussianity from the second and third order terms. During the phase of coherent oscillations around the minimum of the potential, the energy density of the curvaton field for a mode $k$ can be approximated by $\rho_{\sigma} = m^{2}_{\sigma} \sigma_{\rm osc}^2/2$. Repeating the expansion to third order in the $\delta N$ formalism \cite{Sasaki_2006} gives 

\begin{align}\label{s3o}
\z(k) &= \frac{2 \rec}{3} \frac{\sosc'}{\sosc} \d \s_{k}(t_k)  \nonumber \\
&+ \frac{1}{9} \left[ 3 \rec \left( 1 + \frac{\sosc \sosc''}{\sosc'^2}\right) - 4 \rec^2 - 2 \rec^2 \right]\left(\frac{\sosc'}{\sosc}\right)^2 \d \s_{k}^2(t_k)  \nonumber \\
&+ \frac{4}{81} \left[10\rec^4 + 3\rec^5 + \frac{9 \rec}{4} \left(\frac{\sosc^2 \sosc'''}{\sosc'^3} + 3\frac{\sosc''\sosc}{\sosc'^2}\right)  \right. \nonumber \\
&\left. -9 \rec^2 \left( 1 + \frac{\sosc \sosc''}{\sosc'^2}\right) \right] \left(\frac{\sosc'}{\sosc}\right)^3 \d \s_{k}^3(t_k)+ \co(\d \s_{k}^{4})
\end{align}

Note that, the non-linearity parameters $\fnl$ and $\gnl$ are scale dependent. In order to have scale independent parameter one needs $\sigma_{\rm osc}(\s_k)$ not to depend on $\sk$. Straightforward comparison of \eqref{z3o} and \eqref{s3o} gives 

\begin{align} \label{fgdn}
\fnl =& \frac{5}{4} \frac{f_{\rm osc}}{\rec} - \frac{5}{3} - \frac{5}{6}\rec\quad ,  \\
\gnl =& \frac{25}{24} \frac{g_{\rm osc} }{\rec^2} - \frac{25}{6}\frac{f_{\rm osc} }{\rec} - \frac{25}{12}\lp f_{\rm osc} - \frac{10}{9} \rp 
+ \frac{125}{27}\rec + \frac{25}{18}\rec^2  \  , \nonumber 
\end{align}
where 
\begin{align}\label{foscdef}
f_{\rm osc} \equiv 1 + \frac{\sosc \sosc''}{\sosc'^2} &\ , &
g_{\rm osc} \equiv \frac{\sosc^2 \sosc'''}{\sosc'^3} + 3\frac{\sosc''\sosc}{\sosc'^2}\ .
\end{align}

A prime in the above equation indicates derivatives with respect to $\s_k$. The terms proportional to $\rec^{-1}$ show that the faster the curvaton decays, the larger are the non-Gaussianities.  In addition, the lower is the cross-section $\Gamma$, the longer it takes for the system to reach the sudden-decay condition $H \sim \Gamma$. Note that to lower the value of the cross-section means to reduce the magnitude of the curvaton interactions and consequently also their fluctuations. Thus, higher $\rec$ produce smaller magnitude of fluctuations and smaller non-Gaussianities.

The curvaton dynamics is characterized by two distinct regimes. The first is the slow-roll regime of the curvaton, given by

\be \label{srcurv}
3H\dot{\s} + m_{\s}^{2}\s + V_{,\s}^{\rm SI}(\s) \approx 0\ ,
\ee

where $V^{\rm SI}(\s)$ is the self-interacting part of the potential~\footnote{The quadratic mass term emerges only in the small field value limit, hence it is absent in (\ref{srcurv}). However, the Taylor expansion of the total potential V around its minimum makes the mass term reappears in (\ref{eqosc}).}. 

The solution for the slow-roll regime $\s_{SR}$ is a nonlinear function of $\sk$. We assume that it is valid until the time $t_{q}$, when the curvaton reaches its second regime. There, the curvaton oscillates around its quadratic minimum and the self-interactions are no long important. This is know as a coherent oscillating phase, whose dynamics reads

\be \label{eqosc}
\ddot{\s} + 3H\dot{\s} + m_{\s}^{2}\s \approx 0\ .
\ee

The solution for this stage is of the form $\s(t) = \sosc f_{inf}(t)$, where $f_{inf}$ is a function dependent only on the background dynamic given by the inflaton, see ~\cite{Byrnes_Tarrant_2015, Enqvist_2009}. We suppose an  instantaneous transition between the slow-roll and the coherent oscillation regimes and match the respective solutions, which allows us to write $\sosc = \b \s_{\rm SR}(t_q)$, where $\b$ is a constant parameter. Therefore $\sosc$ is proved to be dependent on $\sk$ as well.

We are interested in a particular set of self-interactions curvaton models, where the $\sk$ dependence on $\sosc$ is input by hand in order to better fit observational results. In the next section we show how recent observations suggest the behavior needed for $\fnl$, and, consequently, $\sosc(\sk)$.

\section{Constructing viable models from $\fnl$} \label{SecRecon}
The Planck Collaboration \cite{Planck_2018_6} showed that the strength of the non-Gaussian signal for $\fnl$ does not go beyond order unity, indicating that primordial non-Gaussianities are seemingly very small. A way out of this constraint is to consider scale-dependent non-Gaussianity models, in order to have large values of $\fnl$ away from the observed scales (in particular, away from the pivot scale used for the CMB maps). Models with a change of sign in $\fnl$, that remains close to zero over a limited range of wave-numbers, can be adjusted to satisfy the present observational constraints. Evidently, such range of wave-numbers must be identified with the CMB scales which constrains the free-parameters of the models. This procedure allow us to study how much fine-tuning is required to fit the observational data. The scale-dependent models are particularly interesting when one needs high values of non-Gaussianities, for instance, to account for the CMB anomalies \cite{Byrnes_Tarrant_2015, Byrnes_2016_2, Byrnes_2016}. In the following, we analyze how to construct models with a change of sign in the $\fnl$ parameter and its effects on the dynamics of the curvaton field.

\subsection{Crossing $\fnl$ parameter} \label{sec_cross}
In the literature, the non-Gaussianity parameter $\fnl$ is typically parametrized as a power law given by

\begin{equation}\label{fnl_power}
\fnl(k) =\fnl^0\left(\frac{k}{k_0}\right)^{n_{\fnl}}
\end{equation}

where $\fnl^0$ is the amplitude at a given pivot scale $k_0$ and the index $n_{\fnl}$ is a constant \cite{Sefusatti_2009,Byrnes_2010}. However, this parametrization is no longer valid if $\fnl$ crosses the zero, namely if it changes sign~\cite{Byrnes_Tarrant_2015}. As we present on \ref{SubSec_Scale}, a better suited parametrization allows for multiple changes in sign.

We can find the value of $\sosc$ in terms of $\sigma_k$ where the change of sign happens.  The RHS of \eqref{foscdef} shows that

\ba \label{cond_fnl}
f_{\rm osc} = \frac{\lp \sosc^2 \rp ''}{2 \sosc'^2} = 0 \quad  
\Rightarrow \quad \lp \sosc^2 \rp '' = 0\ .
\ea

The conditions for $f_{\rm osc}$ and $\fnl$ to cross zero are different. Nevertheless, for $\rec \sim 0.05$ the value of the last two terms on the RHS of \eqref{fgdn} are of order unity. The crossing is still guaranteed as long as the scale dependence of $\fnl$ is strong enough. Thus, instead of the crossing of $\fnl$, we consider the condition for $f_{\rm osc}=0$. Equation \eqref{cond_fnl} implies that the crossing is an extremal point for

\be 
\lp \sosc^2 \rp ' = 2 \sosc \sosc' \ .
\ee

The function $\sosc$ is assumed to be a monotonic function of $\sk$ and the derivative $\sosc '$ must not cross zero, otherwise $f_{\rm osc}$ diverges. Moreover, the value of $k$ at the crossing of $f_{\rm osc}$ differs from the extreme of $\sosc '$ due to the factor $\sosc$. As a fact, at the extremal of $\sosc '$ we have $\sosc'' = 0$, which means $\fosc = 1$ instead of $0$. Notwithstanding, it suffices that $\sosc '$ has one extremal point for \eqref{cond_fnl} to be satisfied and, given the scale-dependence of the system, we expect that the value of $k$ for these two conditions should be close.

In resume, the function $\sosc$ is monotonic and $\sosc'$ has an extremal point but it is never zero, hence it is always positive or negative. The function $\sosc''$ does change sign at least once and must vary enough to guarantee that $\fnl$ also changes sign. There are different ways in which one can implement these features. In the next section we show one way to construct the curvaton potential in order to have exactly this kind of behavior. We study the parameter-space of the model by combining the observational data with the conditions on $\sosc$ and its derivatives.

\subsection{Constructing the curvaton potential} \label{sec_recon}

We start by separating the slow-roll regime of the curvaton into two. Near $\s_{\rm SR}(t_q) \equiv \sigma_q$ the curvaton potential is close to quadratic, which persists until the minimum of the potential at the origin $\sigma = 0 $. This guarantees that the results from the conventional self-interaction curvaton scenario (\ref{SubSec_SI_Curvaton}) are valid. Away from  $\sigma_q$, we need to consider the full expression of the potential for the evolution around observable scales $\sk$. Therefore, solving the slow-roll equation for these two regimes gives

\be \label{sr_rec}
\int_{\sk}^{\sq} \frac{\dd \s}{\bar{V}_{, \s}} \sim \int_{\sk}^{\s_m} \frac{\dd \s}{\bar{V}_{, \s}} + \int_{\s_m}^{\sq} \frac{\dd \s}{\s} = -\bar{\eta}_ {\s} \ci\ ,
\ee

where  $\bar{V} \equiv  V/m_{\s}^{2}$, and $\s_m$ is the value when the dynamics changes from slow-roll to quadratic local minimum. We apply matching condition at $\s_m$, and since it is evaluated at the point where both solutions are equal, it cancels from the final expression. Following \cite{Byrnes_Tarrant_2015}, we have also defined

\begin{align}
\ci (t_{q}, t_k) \equiv H_{k}^{2} \int_{t_k}^{t_q} \frac{\dd t}{H(t)} \ , \label{eq_i} \quad 
\bar{{\eta}}_{\s} = \frac{m_{\s}^2}{3 H_{k}^{2}}\ .
\end{align}

Given an appropriate choice of $t_q$, the integral gives $\ci (t_{q}, t_k) \approx 1/\bar{\eta}_{\s}$ during the curvaton slow-roll. Thus, for values of $t << t_q$, the RHS of \eqref{sr_rec} equals $- \bar{{\eta}}_{\s} \ci (t_{q})  \approx -1$.  Close to $\sigma_q$ we have

\be \label{srq}
\int \frac{\dd \s}{\bar{V}_{, \s}} = \int \frac{\dd \s}{\s} = \log \lc \s_q \rc =  \log \lc F(\sk) \rc\ ,
\ee

where $F(\sk)\equiv \sq(\sk)$ is the functional form of $\sq$ in terms of $\sk$. For that we shall use an ansatz that satisfies the conditions imposed on $\fnl$.

Defining the function $G(\s)$ as the primitive of integral containing $\bar{V}_{\s}^{-1}$, we can recast \eqref{sr_rec} as

\begin{align}
G(\s) = \log \lc \sq(\s) \rc + \bar{{\eta}}_{\s} \ci (t_{q}, t)\ .
\end{align}

Note that, when $t = t_q$, the integral above vanishes, and we re-obtain \eqref{srq}. More precisely, if $G(\s)$ is the primitive of \eqref{srq}, we have $G(\sk) = \log \lc \sq(\sk) \rc $. Therefore, we can also rewrite

 \eqref{sr_rec}
as
\be \label{sr_curv_rec}
\int_{\sk}^{\sq} \frac{\dd \s}{\bar{V}_{, \s}} \sim G(\sq) - G(\sk)
\ee
We can invert this equation to find the curvaton potential in terms of the $G(\s)$, namely
\be\label{sr_pot}
V(\s) = m_{\s}^2 \int \frac{\dd \s}{G_{,\s}}\ .
\ee

Therefore, given a physically motivated ansatz for the solution, namely the $F(\sk)\equiv \sq(\sk)$, by integrating \eqref{sr_pot} we recover the potential satisfying the slow-roll dynamics that produces this specific solution. By construction, the slow-roll solution is approximately $G(\s)$. This allows us to compute the curvaton slow-roll parameters and compare it with the observation of the primordial power spectrum. Moreover, using \eqref{fgdn}, we can also compute the non-Gaussianity parameters $\fnl$ and $\gnl$ of the model.

\subsection{Observables from the reconstruction} \label{Obs_from_rec}

Similarly to what we have done for $\fnl$, we can write the derivatives of the potential, namely the slow-roll parameters, in terms of $\sq(\sk)$ and its derivatives. Following (\ref{sr_pot}), we have

\begin{align}
V_{,\s} &= m_{\s}^2\frac{1}{G_{,\s}} = m_{\s}^2 \frac{F(\s)}{F_{,\s}(\s)}; \\
V_{,\s \s} &= - m_{\s}^2 \frac{G_{,\s \s}}{G_{,\s}^2} = m_{\s}^2\left(1 - \frac{F_{,\s \s}(\s) F(\s) }{F_{,\s}(\s)^2}\right)
\end{align}

Comparing the above expression with the definition of $\fosc$, (\ref{foscdef}), one immediately sees that

\begin{align} \label{fosc_ddv}
\fosc + \frac{V_{,\s \s}}{m_{\s}^2} = 2\ .
\end{align}

It is worth remarking that the above expression is independent of the solution $\sq(\sk)$. Since $\fosc = 1$ for $\sosc''(\sk) = 0$, \eqref{fosc_ddv} shows that $ V_{,\s \s} / m_{\s}^2= 1$ at this point as well. We also conclude that

\begin{align}\label{fnl_eta}
\eta_{\s} &= \frac{m_{\s}^2 \left(2 - \fosc\right) }{3 H_{k}^2} 
=  2 \bar{\eta}_{\s} - \frac{4 \rec\,  \bar{\eta}_{\s}}{5}   \left(\fnl + \frac{5}{3} + \frac{5 \rec}{6}\right)\ ,
\end{align}

hence, for any model from our procedure, the parameter $\eta_{\s}$ can be written in terms of $\bar{\eta}_{\s}$ and $\fnl$. Equation \eqref{fnl_eta} generalizes the relation presented in~\cite{Enqvist_2015}, since it is still valid for large values $\fnl$ and $\eta_{\s}$.  We see that in our scenario, there is an additional expression relating $\fnl$, $\eta_{\s}$ and $H_{k}$. Note also that we can recover the condition $\eta_{\s} = 2 \bar{\eta}_{\s}$ from \cite{Byrnes_Tarrant_2015}, if $\fnl(k_0) = -5/3 - 5 \rec/6$.

In section~\ref{sec_cross} we associated the change of sign in $\fosc$ with the second derivative of $\sq''(\sk)$ being zero somewhere along the curvaton trajectory. Now, using \eqref{fosc_ddv}, we conclude that the potential must also have an inflection point, i.e. $V_{,\s \s} = 0$. The inflection point, like for $\fosc$, is not located where $\sq(\sk)'' = 0$.

\subsection{Reconstructing a polynomial potential}

The quartic and higher power polynomial models have been studied in the literature~\cite{Enqvist_2009, Enqvist_2010, Byrnes_2011, Byrnes_Tarrant_2015}. Despite producing scale-dependent non-Gaussianities, these models predict high values of $\eta_{\s}$ over the region of low-$\fnl$. Therefore such models are not favored by the Planck satellite results. Our goal here is to use them only as an example to show how our procedure works. In the next section we shall deal with fitting the model to observations. For a polynomial potential of the form $V(\s) = \frac12 m^2 \s^2 + \lambda \s^n$ with $n > 2$, the curvaton slow-roll solution is given by

\begin{align} \label{sol_poly}
\sq(\sk) = \frac{\sk}{\left[e^{n-2} + (n e^{n-2} - n) \lambda \sk^{n-2}\right]^{1/(n-2)}} \ .
\end{align}

Using \eqref{sol_poly} as the ansatz for the procedure of section~\ref{sec_recon}, we obtain a potential given by

\begin{align}\label{rec_pot}
V(\s) = \frac{1}{2}m^2 \s^2 + \lambda_{eff} \s^n \quad  \mbox{with} \  \lambda_{eff}=\lambda\left( 1-e^{-n+2}\right)\ .
\end{align}

We see that the reconstruction gives a lower value for the coupling constant. The worst case is for $n=3$ where $\lambda_{eff} \approx 0.6321 \lambda$ but already increase to $\lambda_{eff} \approx 0.865 \lambda$ for $n=4$. The higher the power of the self-interaction, more precise is the reconstruction. This kind of shift in our reconstruction procedure does not change the qualitative behavior of our model but it can change some observational scales such as the pivot value at which the non-Gaussianity parameter $\fnl$ changes sign. The important point is that any feature input in the ansatz will also be present at the solutions derived by using the reconstructed potential, hence the consistency of the procedure is guaranteed.

\section{Linearly accelerated models} \label{SecLinear}

In section~\ref{sec_cross}, we described the main features that a solution $\sosc(\sk)$ must have in order to produce a viable curvaton model with a change of sign in $\fnl$. A possible realization of these conditions is the $\sosc''(\sk)$ to be a linear function of $\sk$. Therefore, we consider an ansatz of the form 

\begin{align} \label{ansatz_accel}
\sosc =& a \s +\frac12 b \s^2+ \frac16 c \, \s^3  \quad  ,
\end{align} 

where $a$, $b$ and $c$ are the free parameters of the solution. Applying the construction procedure of the last section, we arrive at the potential

\begin{align} \label{pot_a}
\frac{V(\s)}{m^2} =  V_0 + \frac{b}{3c} \s &+ \frac{\s^2}{6}
-V_{lg}\, \log \left(2 a + 2  b \, \s + c \, \s^2 \right)  \nonumber \\
&+V_{arc}\arctan \left(\frac{b + c \,\s}{\sqrt{-b^2 + 2 a c}} \right)  \quad ,
\end{align}

where the two coefficient $V_{lg}$ and $V_{arc}$ are given in terms of the free parameters as

\begin{align}
V_{lg}=\frac{b^2 - 2 a c}{3 c^2} \quad  &, & V_{arc}=\frac{4 b \left(b^2 - 3 a c\right) }{6 c^2 \sqrt{2 a c-b^2}} \quad .
\end{align}

The argument of the $\arctan$ has the same structure as the ansatz acceleration, i.e. $\sosc'' =  \left(b + c \, \s \right)$ . Therefore, the inflection point for this term happens where the acceleration is zero, $\s_z = -b/c$. However, for the total potential \eqref{pot_a}, the inflection point is shifted away from $\s_z$ due to the presence of the other (sub-leading) terms.

In order to reduce the number of free parameters, and simplify the analysis of the parameter space, we shall fix $a=1/e$, which gives the quadratic curvaton solution in the limit $b = c= 0$.  Note also that we have implicitly assumed $\sosc(0) = 0$. The ansatz is constructed to facilitate the study of the curvaton slow-roll solution and its resulting $\fnl$ parameter. Therefore it is convenient to discuss the model parameter-space in terms of $b$ and $c$ and not in terms of the coefficients of the potential, because the formers are directly connected to the non-Gaussianity parameters $\fnl$ and $\gnl$.

\begin{figure}
	\centering
	\includegraphics[scale=0.45]{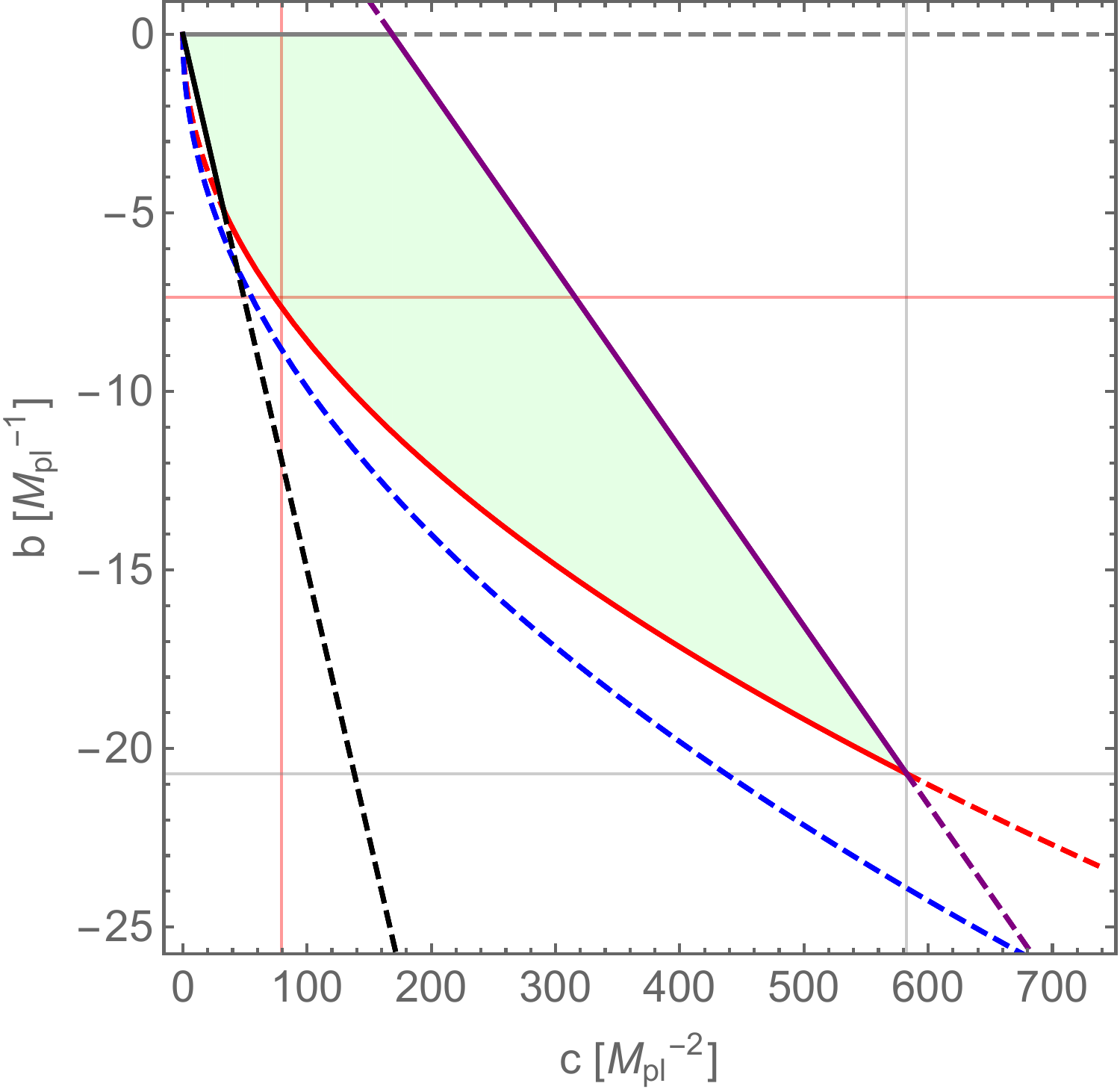}
	\caption{Final allowed region for $b$ in green, when $\smax = 0.15$, $c>0$, including all constraints. The red grid lines the point (b, c) = (-20/e, 216/e) for the model on \ref{Sub_Model}.}
	\label{vinc_bc}
\end{figure}

The first constraint on $b$ and $c$ comes from the change of sign of $\sq''(\sk)$, which should happen during the curvaton slow-roll. Therefore, the point $\s = -b/c$ should be smaller than the initial value of the field $\s_{ini.} \equiv \smax$. That is represented by the black dotted lines in Fig. \ref{vinc_bc}. This also implies that $b$ and $c$ must have opposite signs, inasmuch $\s > 0$ during the slow-roll.

The derivative $\sq'$ should not vanish anywhere, otherwise both $\fnl$ and $\gnl$ diverge. Therefore the models have a positive minimum for $\sq'$, i.e. we must have $b > -\sqrt{2c/e}$ that give the red dashed line constraint in Fig. \ref{vinc_bc}. Note that this condition also avoid divergences in the potential \eqref{pot_a}. Also, $\sq$ is a monotonically increasing function of $\sk$, since its derivative is always positive.

In addition, the $log$ term of \eqref{pot_a} has an argument proportional to $\sosc'$, hence we must also avoid $\sq' = 0$. As a consequence, we must exclude negative values of $c$, i.e $c > 0$. 

The curvaton field should always be positive during the slow-roll, therefore $\sq > 0$, resulting in the blue dotted line in Fig. \ref{vinc_bc}. This condition is, however, less strict and do not contribute since it is always below the red line. 

The last constraint comes from the condition on the curvaton evolution. We want the field to move towards the minimum of the potential at $\s = 0$, hence $\sq$ should never be greater than $\sk$, which gives the purple dash-dotted curve on Fig. \ref{vinc_bc}. To sum up, the system of constraints read

\begin{align}
&\mbox{if}\quad  0 < c < \frac{2}{\smax^2} \quad , \quad - c \, \smax <\, b < 0 \notag\\
&\mbox{if}\quad \frac{2}{\smax^2} < c < \frac{6e - 6}{\smax^2} \quad , \quad -\sqrt{2 c} <\, b < 0 \notag\\
&\mbox{if}\quad \frac{6e - 6}{\smax^2} < c < c_{max} \quad , \quad -\sqrt{2 c} <\, b < \frac{2e-2}{\smax} - \frac13 c \smax\notag\\
&\mbox{where} \quad c_{max} \equiv \frac{3 }{ \smax^2}\left(1 + 2 e + \sqrt{12e - 3}\right)\ .
\end{align}

As a result, we conclude that the higher the value of $\smax$, the smaller is the allowed parameter region for $(b, c)$. In other words, low values of $\smax$ alleviates possible fine-tuning of models.

\subsection{Example A: $b = -20/e$, $c = 216/e$} \label{Sub_Model}

\begin{figure}[h]
	\begin{subfigure}[h]{0.4\textwidth}
		\includegraphics[width=\textwidth]{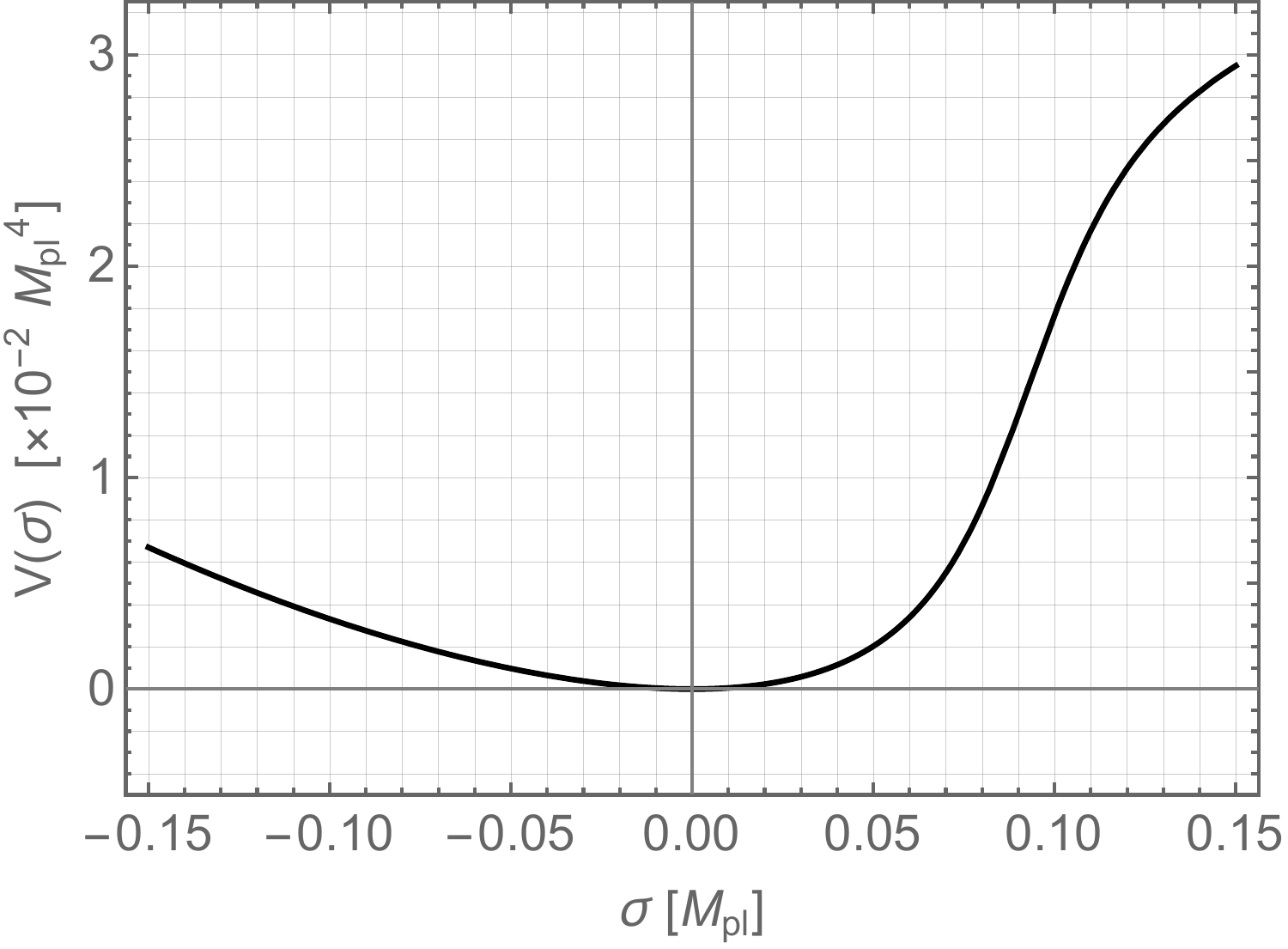}
	\end{subfigure}

\bigskip

	\begin{subfigure}[h]{0.4\textwidth}
		\includegraphics[width=\textwidth]{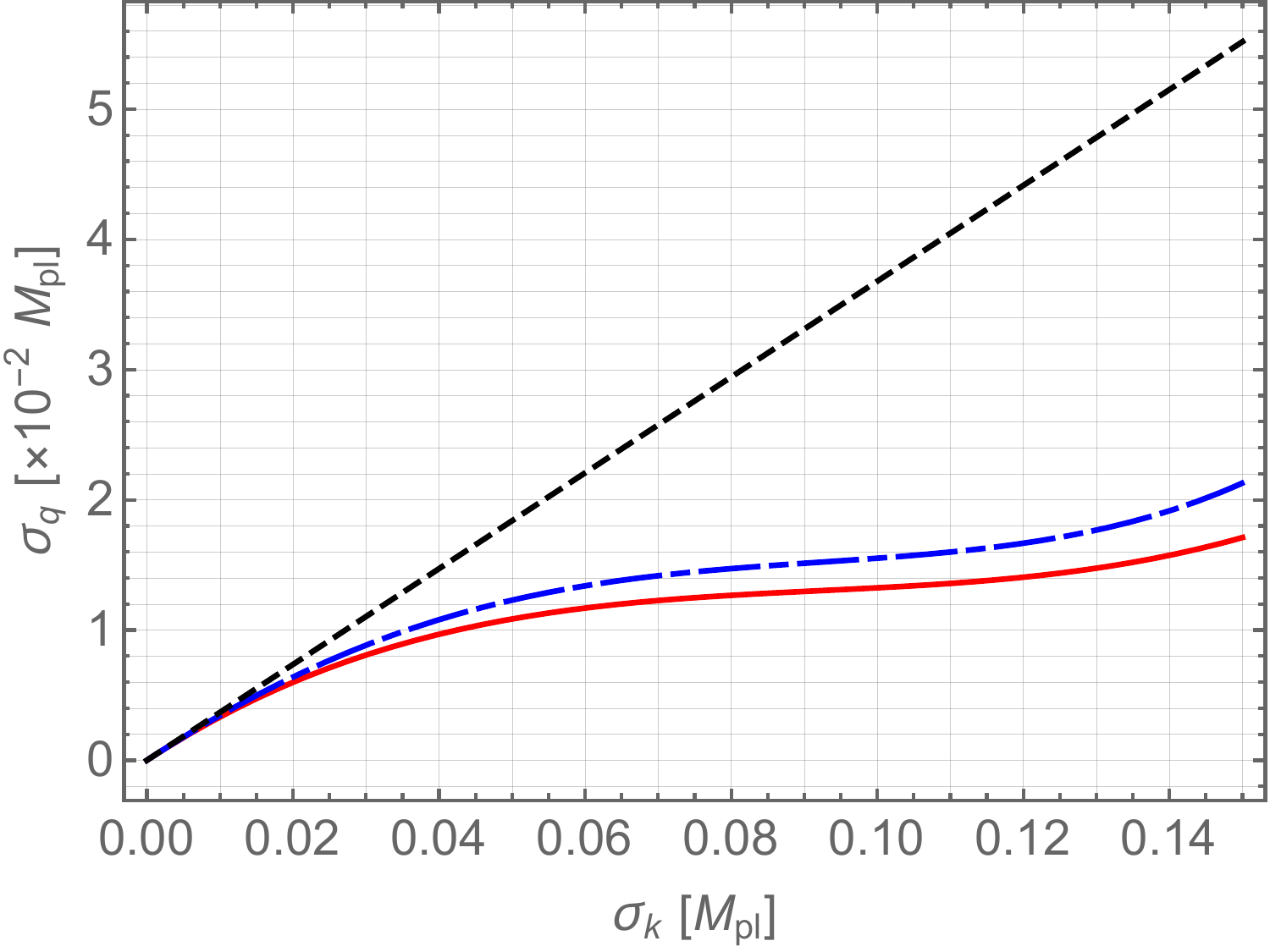}
	\end{subfigure}
	\caption{Top: Curvaton potential $V(\s)$ for the example A, where $b=-20/e$, $c=216/e$. The inflection point is located at $V_{\s \s} = 0$. Bottom: Reconstructed (blue dash-dotted line), Ansatz (red solid line), and quadratic potential (black dashed line) solutions for the curvaton slow-roll equation results in terms of $\sk$.}
	\label{rec_pot_a_and_sol}
\end{figure}

In order to show the behavior of the non-Gaussianities parameter, in this section we study a concrete example by fixing $b = -20/e$, $c = 216/e$. These values are well inside the valid values for $\smax = 0.15$ (see Fig.~\ref{vinc_bc}), and has two interesting properties, i.e. the point $\sq(\sk)'' = 0$ for the ansatz and for the reconstructed solution are close, and lead to reasonable values of $\fnl$ and $\gnl$. Let us first compute the curvaton potential \eqref{pot_a}. To determine the value of $V_0$ we require that $V(0) = 0$, i.e.

\begin{equation}
V_0 \equiv  \frac{2 b \left(3 c-b^2 \right) }{3 c^2 \sqrt{2 c-b^2}}\tan ^{-1}\left(\frac{b}{\sqrt{2 c-b^2}}\right) - \frac{\log 2  }{3 c^2}\left(2 c - b^2\right)  \ 
\label{v0a}
\end{equation}

which gives $V_0\sim 0.016 $ for the chosen values of the parameters. The Taylor expansion of the potential \eqref{pot_a} at $\s = 0$ reads

\begin{align}
V(\s) & = \frac{m^2 \s^2}{2} - b \, \frac{m^2 \s^3}{6} + \co\left(\s^4\right) \quad ,
\label{ansatz_v}
\end{align}

confirming that indeed the potential can be approximated by a quadratic potential close to the origin. In Fig. \ref{rec_pot_a_and_sol} we  show the results of our procedure. The top panel of Fig.~\ref{rec_pot_a_and_sol} displays the potential constructed from the ansatz \eqref{ansatz_accel}, while the in the bottom panel we compare three solutions: the one coming from the reconstructed potential, the original ansatz and the solution for the quadratic potential $V(\s) = m^{2} \s^2/2$.

\begin{figure}[h]	
	\begin{subfigure}[h]{0.42\textwidth}
		\includegraphics[width=\textwidth]{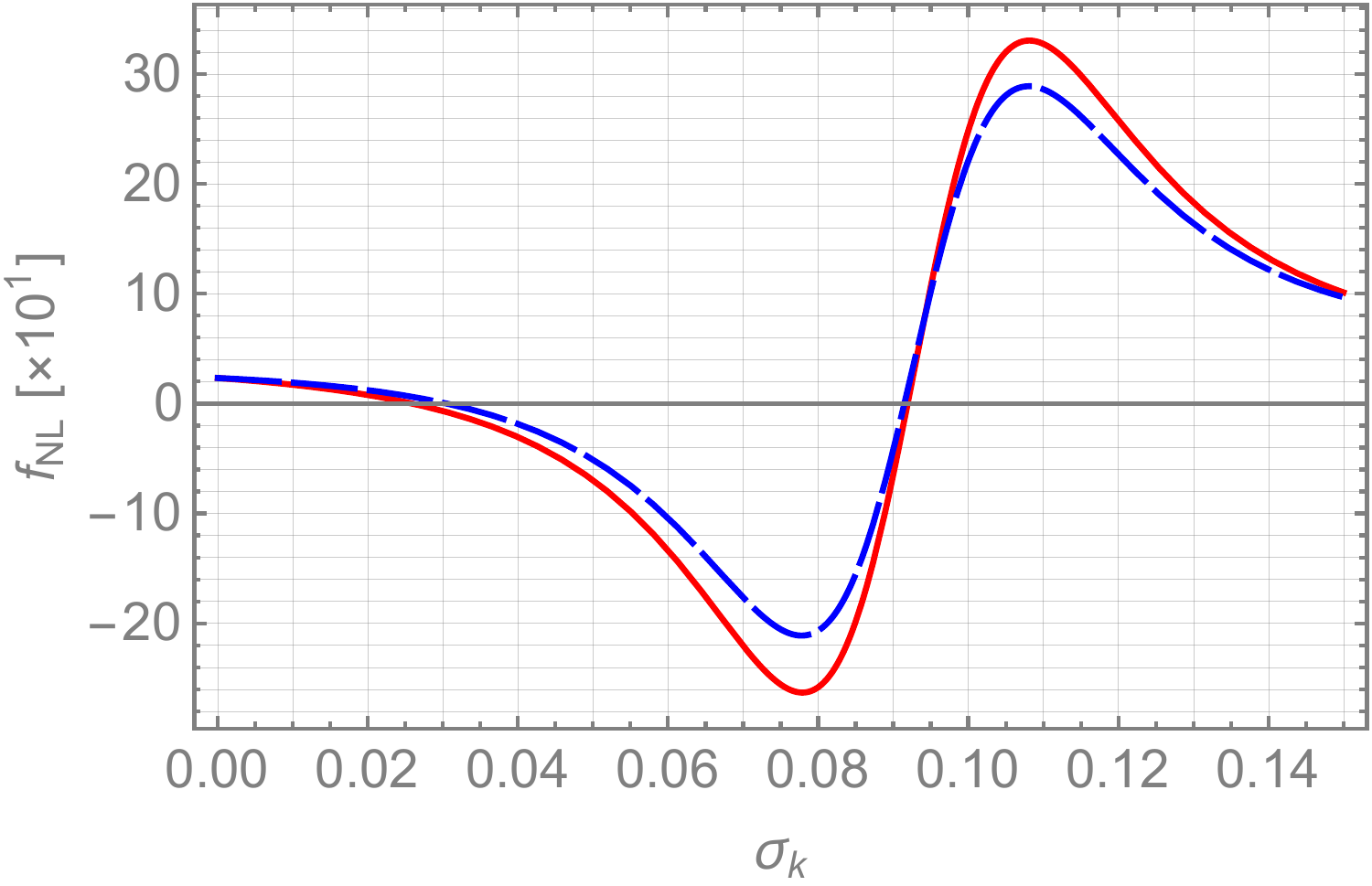}
		\label{fnl_exa}
	\end{subfigure}
	\begin{subfigure}[h]{0.42\textwidth}
		\includegraphics[width=\textwidth]{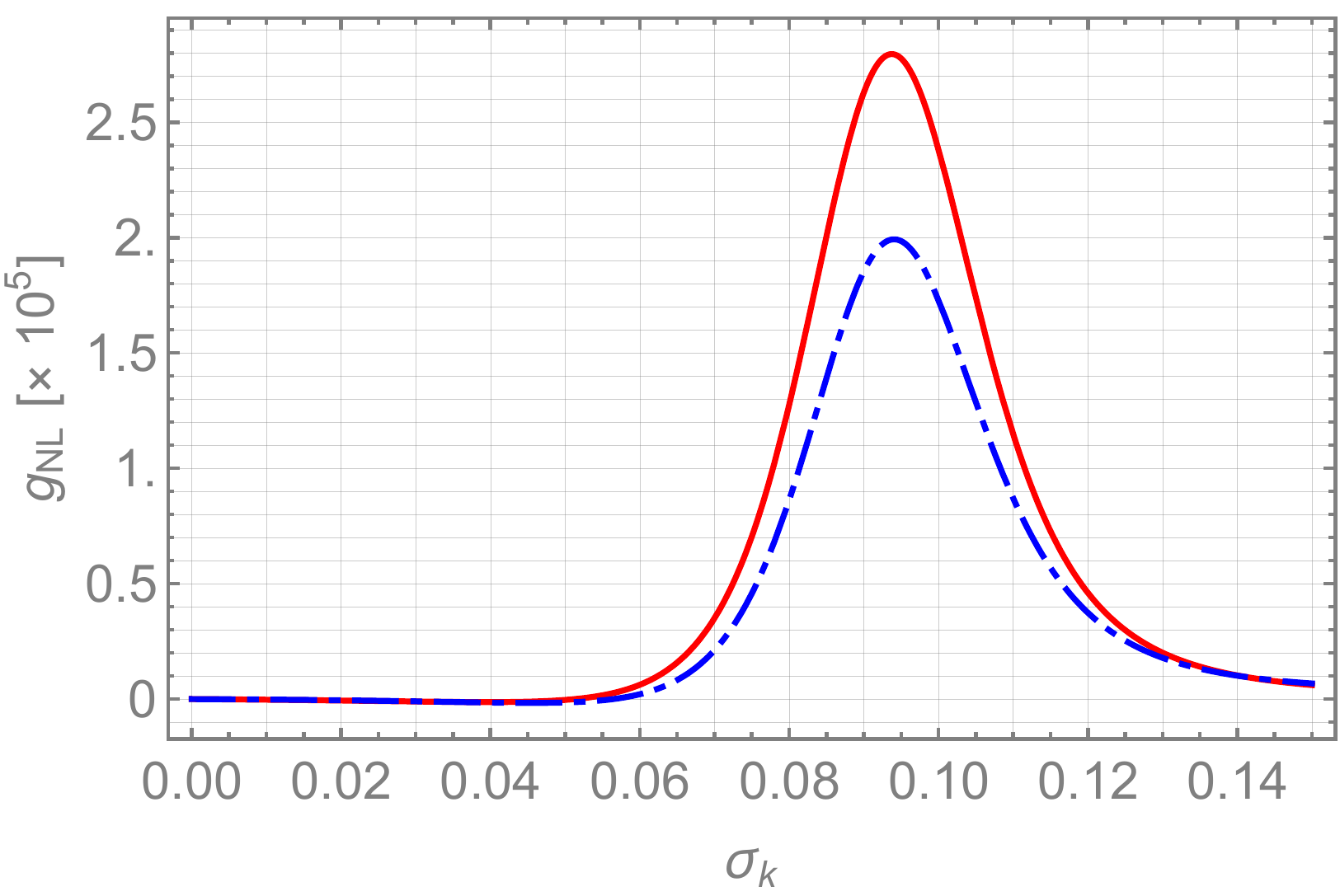}
		\label{gnl_exa}
	\end{subfigure}
	\caption{Top: Reconstructed $\fnl$ (dash-dotted) parameter for example A in comparison to the ansatz (solid) with $b = -20/e$ and $c =216/e$. Below: the same for $\gnl$.}
	\label{fnl_gnl_sigma}
\end{figure}

\begin{figure}[!h]
	\centering
	\includegraphics[scale=0.45]{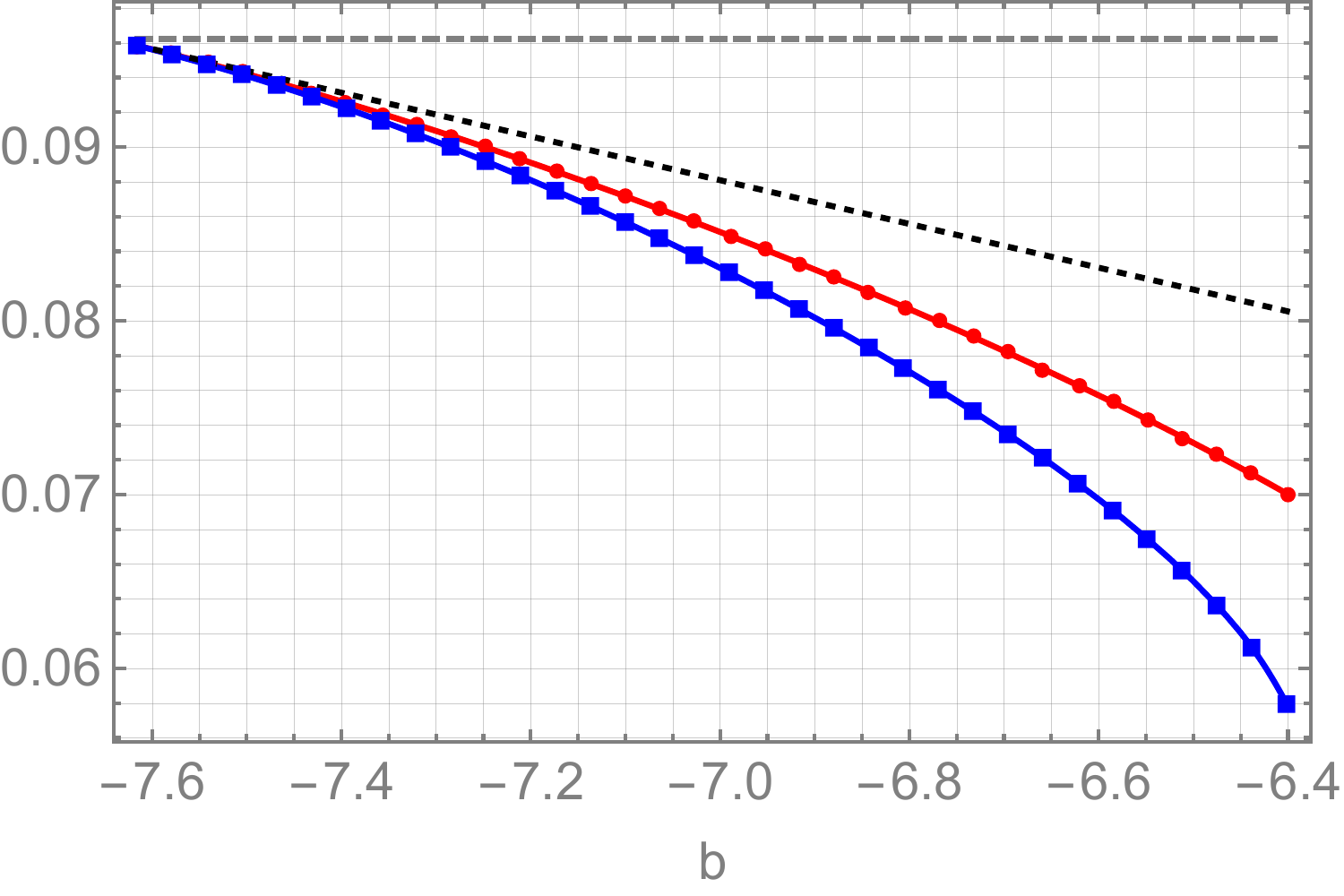}
	\caption{Numerically computed zeros of $\fosc$, as a function of $b$, from ansatz (red, circle) and reconstructed (blue, square) solutions. The ratio $-b/c$ is the black dotted line, while the ratio $b = -\sqrt{2 c/e}$ is the gray dashed line. $c = 216/e$.}
	\label{zeros_fnl_b}
\end{figure}

Note that  the field solution has no maximum or minimum, which guarantees that its velocity is never zero as constructed. The non-Gaussianity parameter $\fnl$ is computed in Fig. \ref{fnl_gnl_sigma}, top panel. The shape of the original ansatz and the reconstructed solution agree but with a small difference in the amplitude. Therefore, we managed to recreate the behavior for $\fnl$ as desired. The agreement between both results grows~\footnote{The results become more alike as the point of their change of sign tends to $-b/c$.} the closer the choice of parameters is to $b = -\sqrt{2 \, c}$, see Fig. \ref{zeros_fnl_b}. However such a choice also implies in stronger non-Gaussianity and running $n_{\fnl}$, beyond the most recent results.

Fig.~\ref{fnl_gnl_sigma} also shows the behavior of the $\gnl$ parameter, bottom panel. It has an extreme at $\sosc'' = 0$, and since the first term of \eqref{foscdef} dominates, it has $\sosc'''$ constant at the extreme. Note that with our choice of parameters, the magnitude of the $\gnl$ is close to the current observational limit, which is  $\gnl \approx 5 \times 10^{5}$.

The authors of \cite{Enqvist_2015} analyze the relation between the features in the curvaton potential and a large running of the scalar spectral index. However, they make no explicit connection between features and the change of sign of $\fnl$ and $\eta_{\s}$, as section \ref{Obs_from_rec}. For our model, recalling \eqref{fnl_eta}, we have

\begin{align}
\eta_{\s} =  \frac{2 m_{\s}^{2}}{3H_{\k}^2} - \frac{4 \rec}{5} \, \bar{\eta}_{\s} \, \left(\fnl + \frac{5}{3} + \frac{5 \rec}{6}\right)
\end{align}

The CMB constrains $\eta_{\s}$ to be of order $10^{-2}$ where $\fnl$ changes sign but the former also depends on the inflationary scale. In turn, to constrain the value of $H_{k}$ we need to evaluate the spectral index and the amplitude of the perturbations (\ref{pert_amplitude}). In fact, we should also include the physics of the reheating~\cite{Kawasaki_2011}. Therefore, in the present analysis we will not fix $H_{k}$. To circumvent this issue, we plot $\fnl$ together to $V_{, \s \s}$. As argued in section~\ref{Obs_from_rec} and also in \cite{Enqvist_2015}, for large values of $\fnl$ we have $\fnl \propto -\eta_{\s}$.

We show that a feature on $V(\s)$ induces a change of sign in $\s''$, i.e. a feature on the solution $\sq(\sk)$. The converse is also true: if we start with an ansatz in which  $\s'' = 0$ somewhere, there will be change of sign in the reconstructed $V_{, \s \s}$. The same goes for $\fnl$. We conclude that features are shared by these different observables.

\begin{figure}[!h]
	\centering
	\includegraphics[scale=0.3]{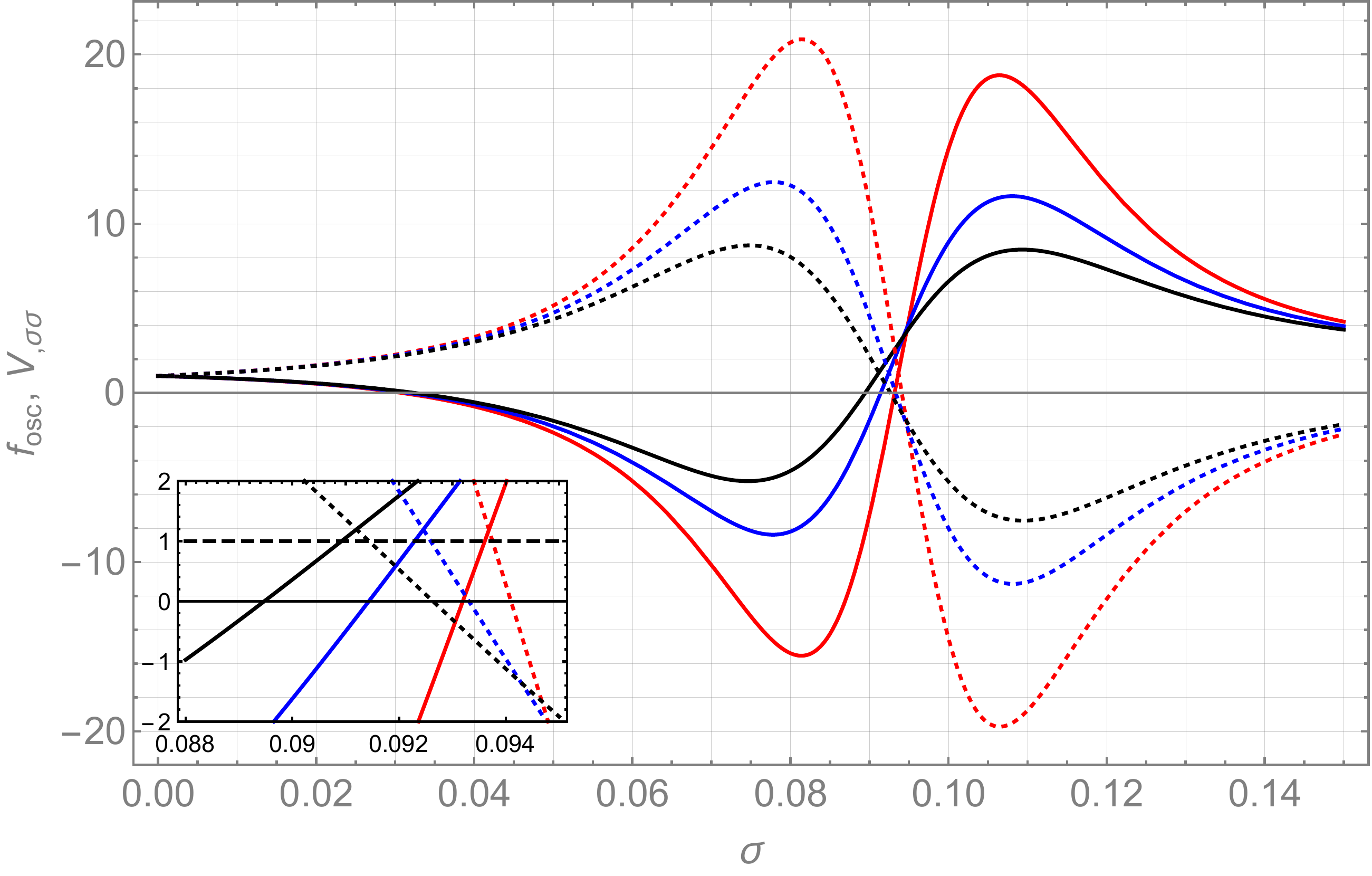}
	\caption{Reconstructed $\fosc$ (solid) and $V_{\s \s}$ (dotted lines) for varying $b$, $a = 1/e$ and $c =216/e$. Their magnitude grow with $|b|$. The distance between the crossing position decreases with growing $|b|$ for the chosen parameter range. In highlight, we show the point where both are equal, which happens for a value a bit above $V_{\s \s} = 1$.}
	\label{eta_fnl_b_comp}
\end{figure}

We have also already demonstrated that the change of sign for the ansatz $\fosc$ and $V_{, \s \s}$ happens in different scales. In \ref{eta_fnl_b_comp} we show that, even for the reconstructed $\fosc$, the zeroes of those functions are symmetric around the point where $\fosc = V_{, \s \s}$. For $\fosc$ the zero is located before $\s = -b/c$, while for the second derivative of the potential it happens after this value. We can also see what is indicated in \eqref{fnl_eta}: when we choose the pivot scale to be where $\fnl = 0$, we have $\eta > 0$.
 
\subsection{Scale dependence effects and CMB Anomalies}\label{SubSec_Scale}

The scale dependence of $\fnl$ can be explicitly written by expanding the integral $\ci \left(t_q,t_k\right)$ (see \ref{eq_i}) in terms of $\log(k/k_0)$. The pivot scale $k_0$ is defined as the value at which $\fnl = 0$. Near the pivot scale, we have

\begin{align}\label{integral_k}
\ci = \log(k/k_0) \left[1 + \epsilon_{0} \log(k/k_0)\right]\quad ,
\end{align}

where $\epsilon_{0}$ is the inflationary first slow-roll parameter at the pivot scale. Note that this modification makes $\fnl$ depend on $\bar{{\eta}}_{\s}$ as well. The reconstructed solution for $\fnl$ and $\gnl$ can then be written in terms of  $\log(k/k_0)$, see Fig.~\ref{fnl_gnl} below. Differently from \ref{fnl_power}, around the pivot scale for our model the $\fnl$ parameter is best described by a $\log$ parametrization, see \cite{Byrnes_Tarrant_2015}.

As we vary the parameters $b$ and $c$, we see that the scale dependence of both non-linear parameters change. Higher values of $|b|$ enhance the non-Gaussianities of the scalar perturbations. On the other hand, higher values of $c$ result in lower values for $\fnl$ and $\gnl$. We illustrate the behavior for variations on $b$ below, Fig. \ref{fnl_gnl}

\begin{figure}[h]	
	\begin{subfigure}[h]{0.45\textwidth}
		\includegraphics[width=\textwidth]{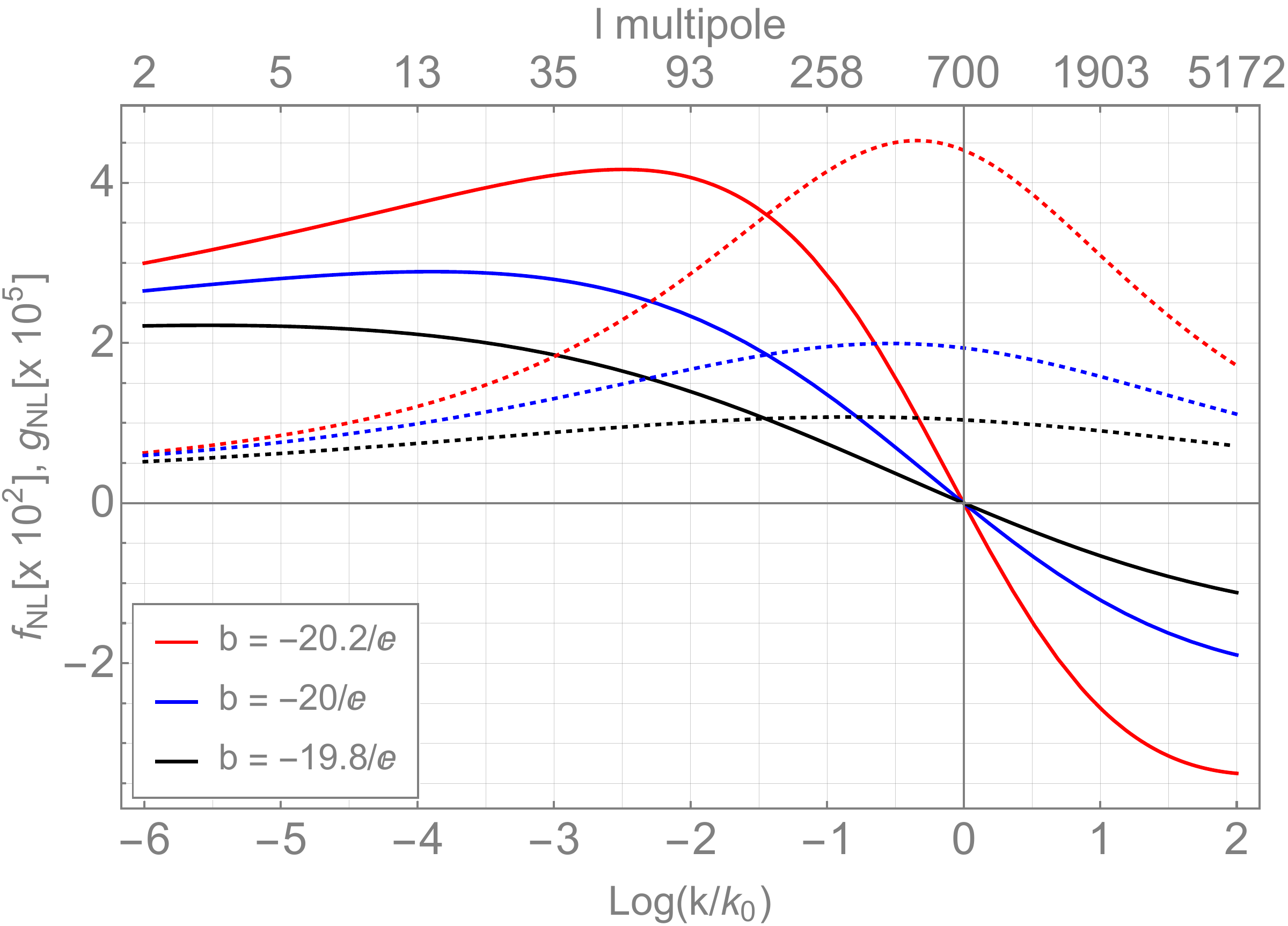}
		\label{fnlgnl_k_multi_b}
	\end{subfigure}
	\caption{non-Gaussianity parameters $\fnl$ (solid) and $\gnl$ (dotted) for the model Example A. We vary $b$ by $\pm 10\%$. The slow-roll parameters were chosen as $\epsilon_{0} = 1/128$ and $\bar{{\eta}}_{\s} = 0.01$. $k_0$ is the pivot scale $0.05 \mathrm{Mpc}^{-1}$, defined as the scale in which $\fnl = 0$.}
	\label{fnl_gnl}
\end{figure}

As it is known, models with scale-dependent non-Gaussianities can account for the CMB anomalies as, for instance, the dipolar modulation \cite{Schmidt_Hui_PRL, Byrnes_Tarrant_2015, Byrnes_2016_2, Adhikari_2016, Adhikari_2018}. Indeed, the model analyzed in~\cite{Schmidt_Hui_PRL, Adhikari_2016} uses the non-Gaussianities to couple short and long scale modes in order to produce the hemispherical asymmetry. The presence of long (super-Hubble) modes of wave-number $k_{l}$ can modulate the Bardeen power spectrum on short scales (inside the horizon). In such a model, the universe remains isotropic, since the dependence on $k$ appears only due to the mode coupling. These models have the advantage, compared to \cite{Byrnes_2016}, that there is no need for large amplitude of the super-Hubble perturbations~\cite{Adhikari_2016}.

The scale dependence of the dipolar modulation roughly follows that of $\fnl$ \cite{Adhikari_2016}. Thus, we expect $\fnl$ to peak at $l < 64$. This provides a new source of observational constraint which helps constraining the parameters of non-Gaussian models.  In Fig. \ref{fnl_gnl} we show the behavior of $\fnl$ for different values of $b$ and $c$. Varying the parameters $b$ and $c$ changes the position of the peak of $\fnl$. Most recent observational results indicate the hemispherical asymmetry to be $A \approx 0.072$ for $l < 64$ \cite{Planck_2015_16, Schwarz_2016}. For shorter scales it reduces to $A < 0.0045$, for $l > 600$\cite{Flender_2013, Quartin_2015}. The region of the parameter space which provides a peak for larger scales is preferred, otherwise the asymmetry would be too high for smaller scales. That is particularly relevant for the quadrupole asymmetry~\cite{Hirata_2009, Kanno_2013, Caballero_2019}. It is also important to note that the spectral index is modulated in scenarios in the the non-Gaussianity is scale-dependent~\cite{Adhikari_2016}, which presents another probe for $\fnl$ and its effects.

Scale-dependent non-Gaussianity can also lead to bias in the cosmological parameter estimation based on the CMB, in special in the presence of scale-dependent trispectrum~\cite{Adhikari_2018}. Depending on the magnitude and scale-dependence of the trispectrum, the bias on the spectral index $n_s$ can reach order of $10^{-2}$, which is of the same order as the expected value of $\eta_{\s}$. Therefore, in different scenarios for non-Gaussian modulation, it is necessary to take into account all effects arising from the  scale-dependence from both bispectrum and trispectrum, in order to rightly access the constraints on the system's parameter space.

So far we have focused on building models in which the reconstruction procedure detailed in Sec. \ref{sec_recon} is well behaved, meaning that the point where $\fnl$ changes sign is the closest possible between the ansatz and reconstructed solution. However, a discordance between both solutions do not mean the choice of parameters is wrong. Such models may not agree with \eqref{ansatz_accel}, but they still provide scale dependence and magnitude for $\fnl$ that fits observational constraints. Therefore, the theoretical predictions from the whole parameter space in Fig. \ref{vinc_bc} should be tested in comparison to observations.

\section{Conclusions}\label{Con}

In this work we built a curvaton model that presents non-Gaussian scalar perturbations, using the $\fnl$ parameter and its dependence on $\sosc(\sk)$ as a guide. Planck latest results indicate that cosmological perturbations at the pivot scale are highly Gaussian,  $\fnlo = -0.9 \pm 5.1$ \cite{Planck_2018_11}. That can be true for either truly Gaussian fluctuations or scale-dependent $\fnl$. Therefore we computed the conditions on $\sosc$ for $\fnl$ to have a change in sign, which respects observational constraints. We used such conditions to build and constrain a curvaton slow-roll solution $\sosc(\sk)$, \eqref{ansatz_accel}. Then, we recovered the curvaton potential that provides this ansatz, computed the reconstructed slow-roll solution and the resulting non-linearity parameters $\fnl$ and $\gnl$.

Scale-dependent non-Gaussianities are also known to be able to produce the hemispherical asymmetry observed in the CMB, in particular via non-Gaussian coupling between scalar modes~\cite{Schmidt_Hui_PRL}. Using the fact that our model predicts a peak in the $\fnl$ parameter for scales larger than the pivot scale, we showed that it is possible to constrain the model using the asymmetry. If the peak is located towards higher values of $l$, constraints on the asymmetry are violated, which shows that the model should not predict a peak for $\fnl$ located at $l > 64$. Future analysis include a precise computation of the asymmetry and additional effects, such as a modulation on the spectral index and a bias on cosmological parameter estimation.

Additional effects are also present when we take into account constraints beyond first and second order scalar perturbations. Despite the fact that we have the amplitude of perturbations, its spectral index (and subsequent running), constraints applied to the reheating scale are also necessary (which is present in $\rec$, since it depends on $\Gamma$ which for the sudden decay approximation will have the same value as $H_{reh.}$.). There are 5 free parameters in the model, two coming from inflation $H_{k}$ and $H_{reh.}$, and the ones coming from the parametrization, $a$, $b$ and $c$.  The value of $H_{k}$ is specially important since it enters in the computation of $\bar{{\eta}}_{\s}$. A numerical analysis is needed in order to precisely constrain the parameter space of the model. Our analytical computations do not consider the reheating process, which can slightly change scales for the crossings, as well as the amplitude of the non-linearity parameters.

\begin{acknowledgments}
We would like to thank Giovanni Marozzi for useful comments. The authors would like to thank and acknowledge financial support from the National Scientific and Technological Research Council (CNPq, Brazil). L. F. G. is supported in part by INFN under the program TAsP (\textit{Theoretical Astroparticle Physics}). This study was financed in part by the Coordenação de Aperfeiçoamento de Pessoal de Nível Superior - Brasil (CAPES) - Finance Code 001.
\end{acknowledgments}

\bibliography{BIBLIOGRAFIA}	
	
\end{document}